\documentclass[letterpaper, 10 pt, conference]{ieeeconf}  

\IEEEoverridecommandlockouts                              

\usepackage[table]{xcolor}
\usepackage{textcomp}
\usepackage{amssymb,amsmath,amsfonts}
\usepackage{graphicx}
\usepackage{tikz}
\usepackage{url}

\usepackage{colortbl}
\usepackage{booktabs}
\usepackage{array}
\usepackage{tabularx}

\newcolumntype{L}[1]{>{\raggedright\let\newline\\\arraybackslash\hspace{0pt}}m{#1}}

\DeclareMathOperator*{\diag}{diag}

\DeclareMathOperator*{\argmin}{argmin}

\newcommand{\norm}[1]{\left\lVert#1\right\rVert}

\makeatletter
\DeclareRobustCommand\onedot{\futurelet\@let@token\@onedot}
\def\@onedot{\ifx\@let@token.\else.\null\fi\xspace}

\usepackage[T1]{fontenc}  
\usepackage[b]{esvect}    

\makeatletter
\newlength\xvec@height%
\newlength\xvec@depth%
\newlength\xvec@width%
\newcommand{\xvec}[2][]{%
  \ifmmode%
    \settoheight{\xvec@height}{$#2$}%
    \settodepth{\xvec@depth}{$#2$}%
    \settowidth{\xvec@width}{$#2$}%
  \else%
    \settoheight{\xvec@height}{#2}%
    \settodepth{\xvec@depth}{#2}%
    \settowidth{\xvec@width}{#2}%
  \fi%
  \def\xvec@arg{#1}%
  \def\xvec@dd{:}%
  \def\xvec@d{.}%
  \raisebox{.2ex}{\raisebox{\xvec@height}{\rlap{%
    \kern.05em
    \begin{tikzpicture}[scale=1]
    \pgfsetroundcap
    \draw (.05em,0)--(\xvec@width-.05em,0);
    \draw (\xvec@width-.05em,0)--(\xvec@width-.15em, .075em);
    \draw (\xvec@width-.05em,0)--(\xvec@width-.15em,-.075em);
    \ifx\xvec@arg\xvec@d%
      \fill(\xvec@width*.45,.5ex) circle (.5pt);%
    \else\ifx\xvec@arg\xvec@dd%
      \fill(\xvec@width*.30,.5ex) circle (.5pt);%
      \fill(\xvec@width*.65,.5ex) circle (.5pt);%
    \fi\fi%
    \end{tikzpicture}%
  }}}%
  #2%
}
\makeatother

\makeatletter
\renewcommand*\env@matrix[1][\arraystretch]{%
  \edef\arraystretch{#1}%
  \hskip -\arraycolsep
  \let\@ifnextchar\new@ifnextchar
  \array{*\c@MaxMatrixCols c}}
\makeatother

\definecolor{commentcolor}{gray}{0.5}
\usepackage{algorithm}
\usepackage{algpseudocode}

\algnewcommand{\LineComment}[1]{\State \textcolor{commentcolor}{\(\triangleright\) #1}}
\algnewcommand{\To}{\textbf{to}}
\algnewcommand{\Break}{\textbf{break}}
\algnewcommand{\Continue}{\textbf{continue}}
\algnewcommand{\IIf}[1]{\State\algorithmicif\ #1\ \algorithmicthen}
\algnewcommand{\EndIIf}{\unskip}
\algnewcommand{\var}[1]{\textit{#1}}
\algnewcommand{\func}[1]{\textsc{#1}}

\newcommand\vA{\mathbf{A}}
\newcommand\vD{\mathbf{D}}
\newcommand\vF{\mathbf{F}}

\newcommand\vI{\mathbf{I}}

\newcommand\vY{\mathbf{Y}}
\newcommand\vZ{\mathbf{Z}}

\newcommand\vu{\mathbf{u}}

\newcommand\vx{\mathbf{x}}
\newcommand\vy{\mathbf{y}}
\newcommand\vz{\mathbf{z}}

\newcommand\vtF{\boldsymbol{\Phi}}
\newcommand\vtG{\boldsymbol{\Gamma}}
\newcommand\vtL{\boldsymbol{\Lambda}}
\newcommand\vtX{\boldsymbol{\Xi}}
\newcommand\vtY{\boldsymbol{\Psi}}
\newcommand\vtf{\boldsymbol{\phi}}
\newcommand\vtvf{\boldsymbol{\varphi}}

\newcommand\vtx{\boldsymbol{\xi}}
\newcommand\vty{\boldsymbol{\psi}}

\newcommand\cD{\mathcal{D}}
\newcommand\cF{\mathcal{F}}

\newcommand\cK{\mathcal{K}}
\newcommand\cL{\mathcal{L}}
\newcommand\cR{\mathcal{R}}

\newcommand\hvx{\hat{\mathbf{x}}}

\newcommand\koop{\cK_t}
\newcommand\vvf{\mathbf{f}}
\newcommand\vvB{\mathbf{B}}

\newcommand\bR{\mathbb{R}}
\newcommand\bX{\mathbb{X}}
\newcommand\bC{\mathbb{C}}

\algrenewcommand\algorithmicrequire{\textbf{Input:}}
\algrenewcommand\algorithmicensure{\textbf{Output:}}

\usepackage{cite}
\usepackage{soul}
\usepackage{environ}
\usepackage{multirow}
\usepackage[table]{xcolor}
\usepackage{url}
\usepackage{color}

\usepackage{xcolor}
\usepackage{mathtools}
\usepackage{siunitx}
\usepackage{adjustbox}
\usepackage{float}
\usepackage{subcaption}
\usepackage[font=small,tableposition=top]{caption}
\usepackage{booktabs}
\usepackage{bm}

\usepackage{pgfplots}
\pgfplotsset{compat=newest} 
\usetikzlibrary{plotmarks}
\usepgfplotslibrary{patchplots}
\usepackage{grffile}
\usepackage{threeparttable}
\usepackage{gensymb}
\usepackage{todonotes}

\title{\LARGE \bf
Learning Koopman Operators with Control Using Bi-level Optimization}

\author{Daning Huang$^{1}$, Muhammad Bayu Prasetyo$^{2}$, Yin Yu$^{1}$, Junyi Geng$^{1}$%
\thanks{$^{1}$ Department of Aerospace Engineering, Pennsylvania State University, University Park, PA, 16802, USA.
{\tt\small \{daning, yzy5368, jgeng\}@psu.edu}}%
\thanks{$^{2}$ Department of Engineering Science and Mechanics, Pennsylvania State University, University Park, PA, 16802, USA.
{\tt\small mbp5652@psu.edu}}
}

\begin{document}

\maketitle
\thispagestyle{empty}
\pagestyle{empty}

\begin{abstract}
The accurate modeling and control of nonlinear dynamical effects are crucial for numerous robotic systems. The Koopman formalism emerges as a valuable tool for linear control design in nonlinear systems within unknown environments.
However, it still remains a challenging task to learn the Koopman operator with control from data, and in particular, the simultaneous identification of the Koopman linear dynamics and the mapping between the physical and Koopman states.
Conventionally, the simultaneous learning of the dynamics and mapping is achieved via single-level optimization based on one-step or multi-step discrete-time predictions, but the learned model may lack model robustness, training efficiency, and/or long-term predictive accuracy.
This paper presents a bi-level optimization framework that jointly learns the Koopman embedding mapping and Koopman dynamics with exact long-term dynamical constraints.
Our formulation allows back-propagation in standard learning framework and the use of state-of-the-art optimizers, yielding more accurate and stable system prediction in long-time horizon over various applications compared to conventional methods.
\end{abstract}

\section{Introduction}

Accurately modeling and controlling nonlinear dynamical effects is critical for robots, especially in challenging scenarios, e.g., aerial robotics~\cite{mousaei2022design}, aerial manipulation tasks~\cite{geng2020cooperative}, offroad driving~\cite{triest2023learning}. These scenarios often exhibit nonlinear effects, such as the coupling between translation and rotational motion, the self-motion and the manipulated objects, or the complex dynamics due to the environment, making control design difficult. Traditional methods, such as state feedback~\cite{mellinger2011minimum} or optimization-based control~\cite{ji2020robust}
, require full knowledge of the system model to predict dynamics and design controllers. However, real-world effects such as wind gusts, boundary layer effects, rough terrain for mobile robots, and hidden dynamics of chaotic nonlinear effects are too complex to be fully captured, leading to poor control performance under these scenarios. As a result, new approaches are needed to model and control these systems accurately and efficiently, especially when faced with complex, uncertain, or rapidly changing environments. 


Data-driven approaches have been successful in capturing unknown dynamics and patterns in complex systems~\cite{schmid2010dynamic, berger2015estimation}, allowing for accurate dynamics prediction. However, in many cases, these methods produce nonlinear models and hence require nonlinear control methods such as iterative Linear Quadratic Regulator (iLQR)~\cite{li2004iterative} or Nonlinear Model Predictive Control (NMPC)~\cite{gros2020linear} to achieve effective system control. These control methods can be computationally expensive as the system states increases, making them infeasible for real-time applications where fast and accurate control is essential. Although some Reinforcement Learning (RL) approaches~\cite{hu2022learning}, either model-based or model-free, can also achieve good performance on nonlinear control, they often suffer from sampling inefficiency and lack of generalizability. Therefore, there is a need for more efficient control methods that can be used in conjunction with data-driven techniques to enable real-time control of nonlinear systems. 

Koopman operator has recently attracted growing interest and shown great potential to provide an elegant way of addressing the control problem under unknown dynamics~\cite{Brunton2016, Mauroy2016, SURANA2016, mauroy2020koopman, bevanda2021koopman, Goswami2022}. It embeds the nonlinear system dynamics in a lifted, higher-dimensional space where the dynamics is governed by a linear but possibly infinite dimensional operator.
Data-driven methods for identifying the Koopman models have gained considerable attention due to the strong expressive power and the rigorous operator-theoretic guarantees~\cite{abraham2019active,Rosenfeld2021, schulze2022data}. The learned linear system on the embedded space is readily amenable for linear or bilinear control techniques. 
Conventionally, the possible representation of the embedding is given as a predefined dictionary~\cite{Goldschmidt2021,Kaiser2021,Klus2022}.
However, finding the mapping between the physical and Koopman states and selecting the embedding representation remains a challenging task, especially in terms of maintaining the predictive accuracy and generalizability.

There are some existing approaches focusing on learning the mapping expressed by deep neural networks~\cite{Li2017,Wang2021}, and then apply linear control methods~\cite{Folkestad2022, shi2022deep}; these work show an improvement in the model compactness as well as predictive accuracy. However, due to the lack of capability to handle constraints in standard learning frameworks, the existing Koopman learning approaches often rely on a single-level unconstrained optimization formulation that attempts to minimize either only one step prediction errors, or multi-step prediction errors, typically with hand-tuned weights for penalty terms.
Such approaches not only require significant amount of effort to tune the penalty term coefficients and loss components during practical implementation, but also suffer from increased computational overhead in backpropagation especially when multi-step prediction errors are optimized. As a result, the existing methods suffer from poor training efficiency, and the learned models may lack robustness to data noise and long-term predictive accuracy.

To overcome the above limitations, this paper proposes a bi-level optimization framework to learn the Koopman operator with control by jointly learning the embedding and the Koopman dynamics.
Specifically, in the inner optimization, we minimize the loss in the Koopman embedding space with exact constraints of long-horizon Koopman dynamics; in the outer optimization, we minimize the reconstruction loss in the original space with the inner optimization serving as constraints.  This formulation removes the need to hand-tune weight parameters and exactly enforces Koopman dynamics during the learning process.
Furthermore, our framework reformulates the Koopman dynamics using an integral form to eliminate the nested backpropagation calculations in the conventional formulations to boost up training efficiency while maintaining the compatibility with standard learning frameworks.
Overall, the framework enforces the reproduction of dynamics over entire trajectory and thus mitigates the issues in data noise and the long-term prediction instability, and holds promise for a more accurate and numerically stable predictive model for control applications.

The paper is organized as follows. Section II presents a brief summary of Koopman operator with control, and the standard learning methods. In Section III, we provide the details in the formulation, analysis, and numerical algorithms of the proposed bi-level optimization framework. In Section IV, we present numerical examples to show the effectiveness of the proposed methodology in terms of training efficiency, predictive accuracy and generalizability.  Finally, we conclude the work and point out possible future directions for further investigation in Section V.

\section{Koopman Theory Preliminary}
\subsection{Basic Formulation}

Koopman Bilinear Form (KBF) \cite{Goswami2022,Jiang2022} provides a means to globally bilinearize a control-affine system of the following form,
\begin{equation}\label{eqn:controlAffine}
    \dot{\vx}=\vvf_0(\vx)+\sum_{i=1}^m\vvf_i(\vx)u_i,\quad \vx(0)=\vx_0
\end{equation}
where $\vx\in\bX\subseteq\bR^r$ is the state vector, $\vu=[u_1\dots u_m]^\top\in\bR^m$ is the input vector, $\vvf_0:\bX\rightarrow\bR^r$ is the system dynamics, and $\vvf_i:\bX\rightarrow\bR^r$ are the control input coupling terms.

In the autonomous case \cite{Mauroy2016}, i.e., when $\vu=0$, the system generates a flow $\vF_t(\vx_0)=\vx(t)$ from an initial condition $\vx_0$.  The continuous time Koopman operator $\koop:\cF\rightarrow\cF$ is an infinite-dimensional linear operator such that
$ 
    \koop g = g\circ\vF_t
$ 
for all $g\in\cF$, where $g:\bX\rightarrow\bC$ is a complex-valued observable function of the state vector $\vx$, $\cF$ is the function space of all possible observables, and $\circ$ denotes function composition.
As a linear operator, $\koop$ admits eigenpairs $(\lambda,\varphi)$ such that
$
    \koop\varphi = \varphi\circ\vF_t = e^{\lambda t}\varphi
$
where $\lambda\in\bC$ and $\varphi\in\cF$ are the Koopman eigenvalue and Koopman eigenfunction, respectively.

The infinitesimal generator of $\koop$ associated with $\vvf_0$, referred to as the Koopman generator, is defined as $L_{\vvf_0}=\lim_{t\rightarrow0}\frac{\koop-I}{t}$, where $I$ is the identity operator, and turns out to be the Lie derivative $L_{\vvf_0}=\vvf_0\cdot\nabla$, with eigenpair $(\lambda,\varphi)$,
$
    \dot{\varphi} = L_{\vvf_0}\varphi = \lambda\varphi
$
Given a set of eigenpairs $\{(\lambda_i,\varphi_i)\}_{i=1}^n$, the \textit{Koopman Canonical Transform} (KCT) \cite{SURANA2016} of the control-affine system \eqref{eqn:controlAffine} is
$
    \dot{\vtvf} = \vtL\vtvf + \sum_{i=1}^mL_{\vvf_i}\vtvf u_i,
$
where $\vtL=\diag([\lambda_1,\cdots,\lambda_n])$, $\vtvf=[\varphi_1,\cdots,\varphi_n]$, and Lie derivatives for the control terms are $L_{\vvf_i}=\vvf_i\cdot\nabla$.

Suppose the set of eigenfunctions is sufficiently large, such that $\vtvf$ span an invariant space for $L_{\vvf_i}$, i.e., each of $L_{\vvf_i}$ can be represented using a $l\times l$ matrix $\vD_i$, $L_{\vvf_i}\vtvf=\vD_i\vtvf$, then the KCT can be brought to a bilinear form \cite{Goswami2022},
$
    \dot{\vtvf} = \vtL\vtvf + \sum_{i=1}^m\vD_i\vtvf u_i.
$
Often it is difficult to directly obtain the eigenfunctions of $\koop$, and instead it is more convenient to learn the bilinear dynamics in a lifted coordinates via a mapping $\vz=\vtf(\vx)\in\bR^n$, e.g., parametrized by a neural network, leading to the commonly used Koopman Bilinear Form (KBF) \cite{Goswami2022,Jiang2022},
\begin{equation}\label{eqn_kbf}
    \dot{\vz}=\vA\vz+\sum_{i=1}^m\vvB_i\vz u_i.
\end{equation}
The eigendecomposition $\vA=\vtF\vtL\vtY^H$ reproduces the Koopman eigenvalues $\vtL$ and the Koopman eigenfunctions $\vtvf=\vtY^H\vz$, where $\Box^H$ denotes conjugate transpose.  The original states are recovered from an inverse mapping $\vx=\vtf^{-1}(\vz)\equiv\vty(\vz)$.
In standard KBF \cite{SURANA2016}, $\vtf^{-1}$ is a linear mapping, but nonlinear versions of $\vtf^{-1}$ have also been proposed, e.g., in \cite{Lusch2018,schulze2022data}.

\subsection{Learning of KBF Model}
The KBF model is usually identified from data, that is typically obtained at finite sampling rate.  Suppose there are $K$ sampled trajectories with inputs, each having $N+1$ steps; denote the dataset $\cD=\{(\hvx_i^{(k)},\vu_i^{(k)})_{i=0}^N\}_{k=1}^K$.  The parameters to be learned from $\cD$ include the system matrices
$$
\vtG=[\vA,\vvB_1,\vvB_2,\cdots,\vvB_m]
$$
and the encoder $\vtf$ and decoder $\vty$.

A brutal force formulation to learn KBF is an ODE-constrained optimization problem.  Without loss of generalizability, we show the case for one trajectory data.
\begin{equation}\label{eqn_node}
\min_{\vtG,\vtf,\vty} \cL_e(\vz;\vtf,\cD) + \cL_d(\vz;\vty,\cD) + \cL_r(\vtf,\vty;\cD),\ \mathrm{s.t.}\ \eqref{eqn_kbf}
\end{equation}
where three loss terms are introduced:
(1) Encoder loss:
$
\cL_e = \frac{1}{(N+1)n}\sum_{i=0}^N \norm{\vtf(\hvx_i)-\vz_i}^2
$,
(2) Decoder loss:
$
\cL_d = \frac{1}{(N+1)r}\sum_{i=0}^N \norm{\hvx_i-\vty(\vz_i)}^2
$,
(3) Reconstruction loss:
$
\cL_r = \frac{1}{(N+1)r}\sum_{i=0}^N \norm{\hvx_i-\vty(\vtf(\hvx_i))}^2
$.
The term $\cL_e=0$ is a necessary condition for satisfying the invariant space assumption in KBF formulation, $\cL_d$ penalizes on the $N$-step prediction error at each of $N$ steps, and $\cL_r$ penalizes on the reconstruction error.
In addition, sometimes a regularization term $\cR(\cD)$ is included to improve the model generalizability.

However, due to the ODE constraint, optimizing problem \eqref{eqn_node} requires an adjoint ODE solver, that is expensive to evaluate; also a generic adjoint solver does not leverage the bilinear structure of KBF for training efficiency.

A more widely-used approach is to discretize the KBF model \eqref{eqn_kbf} with a time step size $\Delta t$.  Assuming a zeroth-order hold of input $\vu_k$ at time $t_k$ and a sufficiently small $\Delta t$, discrete-time KBF takes the following form of up to first-order time accuracy \cite{Peitz2020}
\begin{equation}\label{eqn_dkbf}
    \vz_{k+1} = \vA_d\vz_k + \sum_{i=1}^m\vvB_{d,i} \vz_k u_{k,i} \equiv \vA_k\vz_k
\end{equation}
where $\vA_d=\vI+\vA\Delta t$, $\vvB_{d,i}=\vvB_i \Delta t$, and $\vA_k=\vA_d + \sum_{i=1}^m\vvB_{d,i} u_{k,i}$ is effectively a family of matrices parametrized by $\vtG$.  Subsequently, the optimization is reduced to an equality-constrained formulation,
\begin{subequations}\label{eqn_dopt}
\begin{align}
    \argmin_{\vtG,\vtf,\vty}&\ \cL_e+\cL_d+\cL_r \\
    \mathrm{s.t.} &\ \vz_k = \vA_k\vz_{k-1},\quad k=1,\cdots,N \\\label{eqn_dopt_c}
    &\ \vz_0 = \vtf(\hvx_0)
\end{align}
\end{subequations}
which will be referred to as \textit{single-level optimization} (SLO).

Note that the equality-constrained form of SLO is chosen for the sake of clarity, and it is equivalent to the more commonly used unconstrained formulation (e.g., \cite{Li2017,Lusch2018,Wang2021,Folkestad2022}), since the $N+1$ intermediate variables $\{\vz_k\}_{k=0}^N$ can be solved exactly from the $N+1$ constraints, starting from \eqref{eqn_dopt_c}.
Particularly, when $N=1$, the SLO reduces to a {single-step formulation} \cite{Li2017,Wang2021,Folkestad2022}, with encoder loss
$
\cL_e=\frac{1}{Nn}\sum_{k=1}^N \norm{\vtf(\hvx_k)-\vA_{k-1}\vtf(\hvx_{k-1})}^2
$, and decoder loss
$
\cL_d=\frac{1}{Nr}\sum_{k=1}^N \norm{\hvx_k-\vty(\vA_{k-1}\vtf(\hvx_{k-1}))}^2
$.


For the learning of KBF, the SLO formulation poses three potential concerns.
\textit{First}, the time discretization of the learned KBF model is first-order time accurate; this may result in the error accumulation that impairs predictive accuracy over long time horizon, and may also pose a challenge when learning dynamics sampled {at low frequency}.
\textit{Second}, for mathematical rigor, the validity of KBF hinges on the satisfaction of $\cL_e=0$ for the invariance of Koopman subspace.  However, in the sum of losses, upon convergence of the model training, each loss term would typically reach a small but nonzero value, and a nonzero $\cL_e$ indicates an inaccurate KBF operator that produces error in the time horizon of $N$ considered in the training.  During the prediction, the error may start to accumulate within a short time horizon, and limit the long-term predictive capability.
\textit{Third}, while the SLO can be written in an unconstrained form and compatible with common deep learning frameworks, the dynamics losses have a recursive formulation that involves nested evaluation of the intermediate variables $\vz$, leading to a high computational cost in the backpropagation during training on the order of $O(N^2)$, i.e., a quadratic (or at least superlinear) growth with respect to the length of time horizon; this renders the learning process inefficient.

\section{Learning Koopman Operator Using Bi-level Optimization}
We present a new \textit{bi-level optimization} (BLO) to resolve the potential issues of SLO.

\subsection{Integral formulation}

First, to minimize the model error due to time discretization, we employ a general formulation based on numerical integration.  Given a control input $\vu(t)$, integrate on both sides of \eqref{eqn_kbf} over $[t_0, t_N]$,
\begin{align}\nonumber
    \int_{t_0}^{t_N} \dot{\vz} dt &= \int_{t_0}^{t_N} \vA\vz+\sum_{i=1}^m\vvB_i\vz u_i(t) dt \\
    \label{eqn_dsc}
    \Rightarrow\quad \vz_N &= \vz_0 + \sum_{i=0}^{N} w_i\vA_i\vz_i,
\end{align}
where $w_i$ are the weights for numerical integration using evenly spaced data points, which are obtained using the standard composite Newton-Cotes formulae, e.g., trapezoid rule for first-order accuracy and Simpson's 3/8 rule for third-order accuracy.
When $k=1$ and $[w_0,w_1]=[\Delta t,0]$, \eqref{eqn_dsc} reduces to the zeroth-order hold formulation.

Next, using the integral form, SLO simplifies to
\begin{subequations}\label{eqn_iopt}
\begin{align}
    \argmin_{\vtG,\vtf,\vty}&\ \cL_e^b+\cL_d^b+\cL_r \\\nonumber
    \mathrm{s.t.} &\ \eqref{eqn_dsc} \\\label{eqn_iopt_c}
    &\ \vz_k = \vtf(\hvx_k),\quad k=0,\cdots,N-1
\end{align}
\end{subequations}
with new encoder and decoder losses: 
$
\cL_e^b=\frac{1}{Nn} \norm{\vtf(\hvx_N)-\vz_N}^2
$ and
$
\cL_d^b=\frac{1}{Nr} \norm{\hvx_N-\vty(\vz_N)}^2
$.  
The new single dynamics constraint \eqref{eqn_dsc} may appear ``weaker'' when compared to the explicit multi-step dynamics constraints in the standard SLO formulation, but it more accurately represents the long-horizon KBF dynamics at the order of numerical integration.


\subsection{Bi-level optimization}

Next, to mitigate the second issue of SLO, \eqref{eqn_iopt} is reformulated in a BLO form,
\begin{subequations}\label{eqn_bopt}
\begin{align}\label{eqn_bopt_a}
    \argmin_{\vtf,\vty}&\ \cL_e^b+\cL_r \\\label{eqn_bopt_b}
    \mathrm{s.t.} &\ \min_{\vtG} \cL_e^b \\\label{eqn_bopt_c}
    \ &\ \mathrm{s.t.}\ \eqref{eqn_dsc},\ \eqref{eqn_iopt_c}
\end{align}
\end{subequations}
The inner optimization solely minimizes the encoder loss with respect to the system matrices $\vtG$, and would achieve at least the approximate satisfaction of the invariance of Koopman subspace, meaning an accurate enforcement of Koopman dynamics over time horizon of length $N$.  The outer optimization minimizes the losses with respect to the autoencoder $\{\vtf,\vty\}$, while {the decoder loss $\cL_d^b$ is removed}; the argument is that, when the reconstruction loss $\cL_r$ is minimized, the fact that the encoder loss $\cL_e^b$ is always minimized would imply a sufficiently low decoder loss.


\subsection{Algorithm for solving bi-level optimization}
Subsequently, we present an algorithm for solving BLO at a cost of $O(N)$ to resolve the third issue of SLO.

The inner optimization \eqref{eqn_bopt_b}-\eqref{eqn_bopt_c} is converted to an unconstrained one,
$
\min_{\vtG} \norm{\Delta\vz - \vtG\vtx}^2
$, where $\vz_i=\vtf(\hvx_i)$, $\Delta\vz=\vz_N-\vz_0$, $\vtx = \sum_{i=0}^N w_i \vy_i$, and
$
\vy = [\vz^\top, u_1\vz^\top, u_2\vz^\top, \dots, u_m\vz^\top]^\top
$.

For a dataset of $K$ trajectories, define $\vZ=[\Delta\vz^{(1)}, \Delta\vz^{(2)},\cdots, \Delta\vz^{(K)}]$ and $\vtX=[\vtx^{(1)}, \vtx^{(2)},\cdots, \vtx^{(K)}]$, and the unconstrained BLO formulation is
\begin{subequations}\label{eqn_uopt}
\begin{align}\label{eqn_uopt_out}
    \argmin_{\vtf,\vty}&\ \cL_e^K(\vtG,\vtf,\vty)+\cL_r^K(\vtf,\vty) \\\label{eqn_uopt_inn}
    \mathrm{s.t.} &\ \min_{\vtG} \cL_e^K(\vtG,\vtf,\vty)
\end{align}
\end{subequations}
where
\begin{gather*}
\cL_e^K=\frac{1}{K(N+1)n}\norm{\vZ-\vtG\vtX}^2 \\
\cL_r^K=\frac{1}{K(N+1)r}\sum_{i,j}\norm{\hvx_i^{(j)}-\vty(\vtf(\hvx_i^{(j)}))}^2
\end{gather*}

Despite its complexity of two levels, the BLO problem may be solved at relative ease leveraging the coordinate descent strategy.  This is because the system matrices $\vtG$ are only solved in the inner optimization \eqref{eqn_uopt_inn}, and kept fixed in the outer optimization \eqref{eqn_uopt_out}; vice versa for the autoencoder parameters $\vtf,\vty$.  At the inner level, due to its simple quadratic structure, \eqref{eqn_uopt_inn} has a closed-form solution $\vtG = \vtX\vY^+$, where $\Box^+$ denotes pseudo-inverse.  At the outer level, one may employ a typical learning framework based on stochastic gradient descent (SGD) methods, e.g., RMSProp or Adam, to minimize the loss.  A similar strategy is employed in~\cite{Li2017,Wang2021}, where a one-step discrete-time formulation was used.

\begin{algorithm}
\caption{Learning algorithm for Koopman with control based on bi-level optimization}\label{alg:blo}
\begin{algorithmic}[1]
\Require Dataset of $K$ trajectories, length of time horizon $N$, number of epoches $N_{ep}$, number of batches $N_{b}$ for SGD, order of numerical integration $P$.
\Ensure Model parameters $\{\vtG,\vtf,\vty\}$.
\State Initialize $\{\vtG^{(1)},\vtf^{(1)},\vty^{(1)}\}$
\For{$n=1,2,\cdots,N_{ep}$}
\State Form matrices $\vZ$ and $\vtX$ from all data using order $P$.  \Comment{$O(\nu NK)$}
\State Fix $\{\vtf^{(n)},\vty^{(n)}\}$, solve \eqref{eqn_uopt_inn} for $\vtG^{(n+1)}$.  \Comment{$O(\nu^2K)$}
\State Shuffle and split the data into $N_b$ batches.
\For{$i=1,2,\cdots,N_{b}$}
    \State Form matrices $\vZ^i$ and $\vtX^i$ from the $i$th data batch using order $P$.  \Comment{$O(\nu NK/N_b)$}
    \State Fix $\vtG^{(n+1)}$, and update $\{\vtf,\vty\}$ using $\vZ^i$ and $\vtX^i$.  \Comment{$O(NK/N_b)$}
\EndFor
\EndFor
\end{algorithmic}
\end{algorithm}

\subsection{Computational complexity analysis}\label{sec_comp}
The complete Koopman learning algorithm based on BLO is listed in Alg. \ref{alg:blo} and the computational cost of each major step is labelled.  The details are discussed further as follows.

Let $\mathrm{dim}(\vtx)=\nu=n(m+1)$, and typically the number of trajectories $K\gg\nu$.  At the inner level, forming matrices $\vZ$ and $\vtX$ costs $O(\nu NK)$, and solving the least squares problem using SVD costs $O(\nu^2 K)$.  At the outer level, the cost in each batch is dominated by forming the matrices, and in total the cost is $O(\nu NK)$.  Therefore, the computational complexity of BLO is $O(\nu(\nu+N) K N_{ep})$ and scales \textit{linearly} with the length of time horizon; this is in sharp contrast with the quadratic (or at least superlinear) growth in SLO.

\section{Numerical Simulation}
To demonstrate the effectiveness of the proposed approach, we investigate two example nonlinear systems: a well-studied two-dimensional nonlinear system in nonlinear control; and a lightly-damped double pendulum system with dimension 4, which shows that the proposed algorithm can generalize to higher-dimensional systems.

\subsection{A two-dimension nonlinear system}
\subsubsection{Problem setup}
We consider a variant of a well-known nonlinear system~\cite{Brunton2016}:
\begin{subequations}
\begin{align}\label{parabolic}
    \dot{x}_1 &= \mu x_1 + u_1 + u_3 x_1 \\
    \dot{x}_2 &= \lambda(x_2 - x_1^2) + u_2
\end{align}
\end{subequations}
where $\mu=-3$ and $\lambda=-2$ are pre-defined system parameters controlling characteristic time scales, and $(u_1, u_2, u_3)$ are time-varying controls to the system.  The system has an isolated equilibrium point at $\vx_e=(\frac{-u_1}{\mu+u_3}, \frac{u_1^2}{(\mu+u_3)^2}-\frac{u_2}{\lambda})$, and increasing $u_3$ slows down the convergence to $x_{e,1}$.

The trajectories for model training and validation were generated by uniformly sampling initial conditions $\vx_0 \in [-5, 5]\times[-5, 5]$ with 32 points in both directions, with step inputs as control that are randomly generated with $u_i \in [-1.8, 1.8]$.  Another 100 trajectories are randomly sampled from the same range of $\vx_0$ and $u_i$.  All trajectories were generated using by 4th order Runge Kutta with a time step size of 0.08s for 25 steps. All trajectories were normalized to $[0, 1]$.
During training, when a time horizon of $N$ is used, using a sliding window of $N+1$, a $L$-step trajectory can produce $K=L-N$ trajectories for training.

A 4-dimensional embedding space is selected based on empirical observation, where three dimensions are learned using a neural network for encoding/decoding, and the remaining dimension is set to be 1 for the completeness of the basis.
Based on a cross-validation study, both the encoder and decoder have 2 hidden layers with Swish activation and have sizes (16, 16).  For all benchmark cases, the model is trained for 800 epochs with 16 batches using the Adam optimizer, and learning rates of 0.001 and 0.0001 are used for SLO and BLO, respectively.  For consistency in comparison, all algorithms are implemented using the JAX package.

\subsubsection{Results}
\begin{figure}
     \centering
     \begin{subfigure}[b]{0.48\linewidth}
         \centering
         \includegraphics[width=\textwidth]{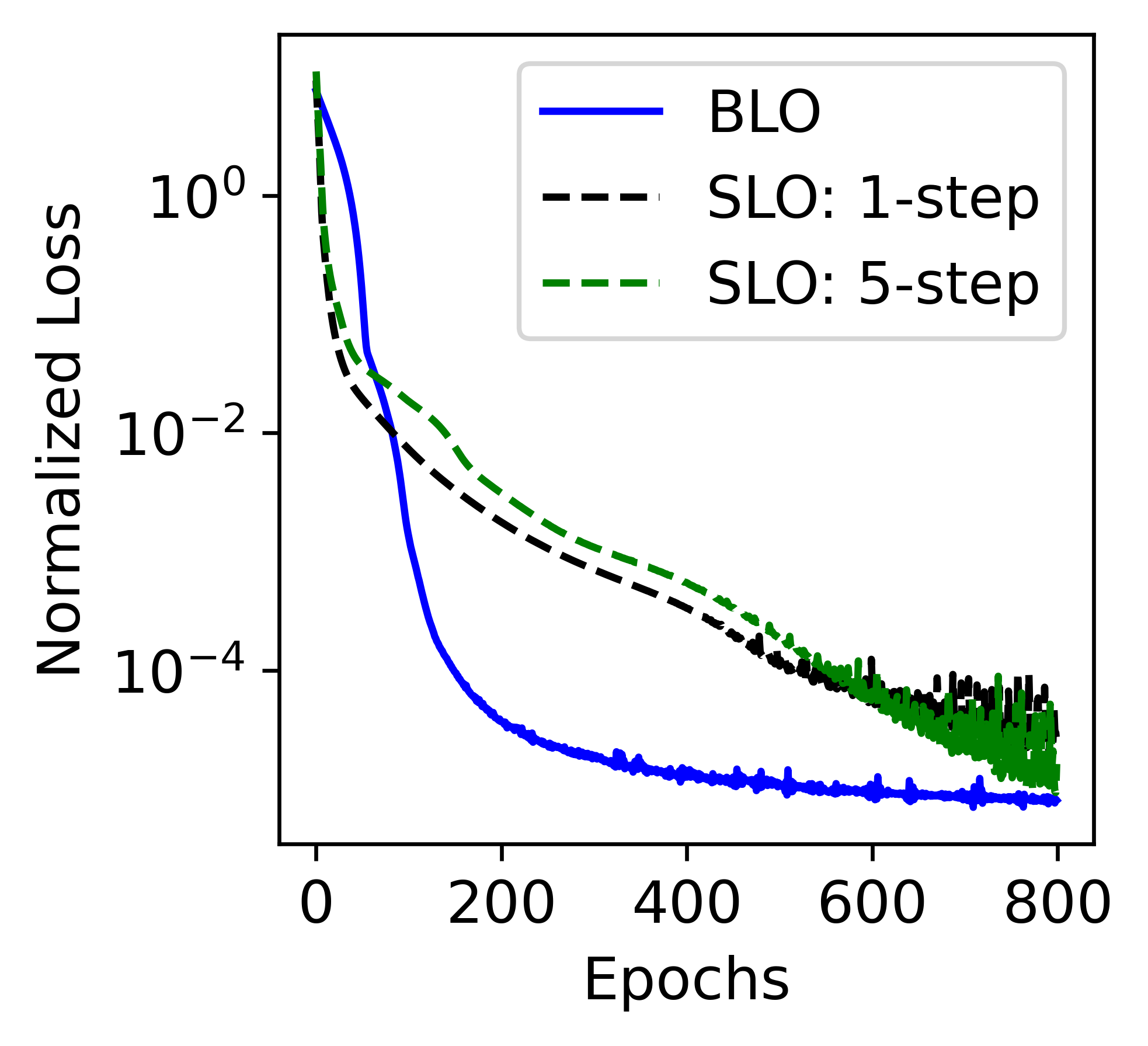}
         \vspace{-5mm}
         \caption{Convergence history}
         \label{fig_basel_h}
     \end{subfigure}
     \begin{subfigure}[b]{0.48\linewidth}
         \centering
         \includegraphics[width=\textwidth]{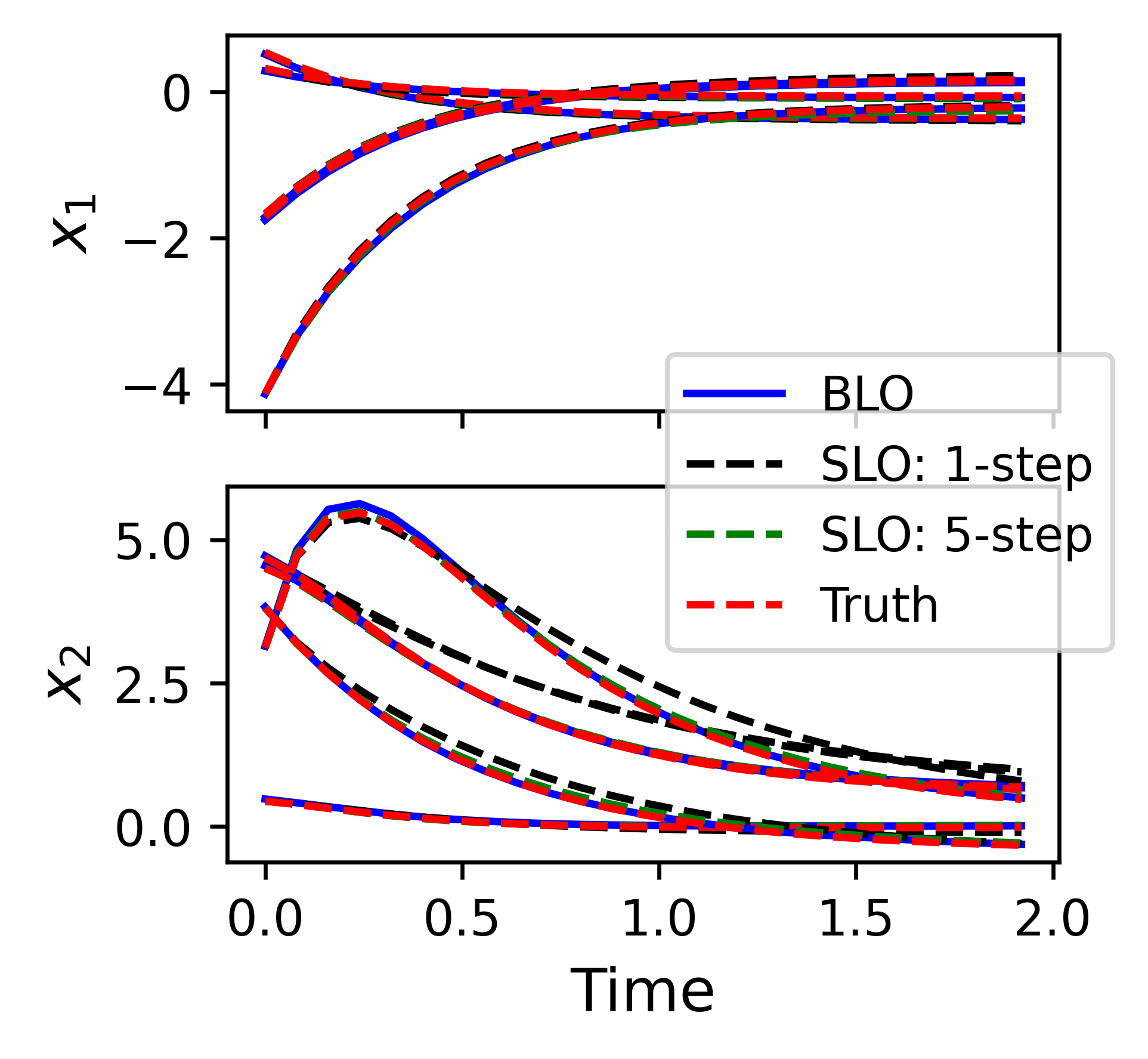}
         \vspace{-5mm}
         \caption{Samples of predicted trajectory}
         \label{fig_basel_p}
     \end{subfigure}
        \caption{Baseline performances of BLO and SLO.}
        \label{fig_basel}
    \vspace{-5mm}
\end{figure}

First, we compare the BLO against a single-step SLO and a 5-step SLO case in Fig. \ref{fig_basel} in terms of convergence characteristics and prediction accuracy.  
Due to the differences in the implementation, only the prediction losses in the original state space are reported, and the losses are normalized by their respective initial values, so that the relative decreases in the loss are compared.
The SLO cases have a similar initial convergence rate and start to show difficulty to reduce the predictive loss further, presumably due to the competing effects with the other losses.  The BLO shows an initially slower convergence rate but achieves convergence within 400 epochs.  In model prediction, the BLO model achieves the lowest error of 3.5\%, which is attributed to the more accurate model representation.  Among the SLO formulation, multi-step case (3.7\%) is relatively better than the single-step case (19\%), as the former accounts for longer horizon in the training to achieve better accuracy.

\begin{figure}
     \centering
     \begin{subfigure}[b]{0.48\linewidth}
         \centering
         \includegraphics[width=\textwidth]{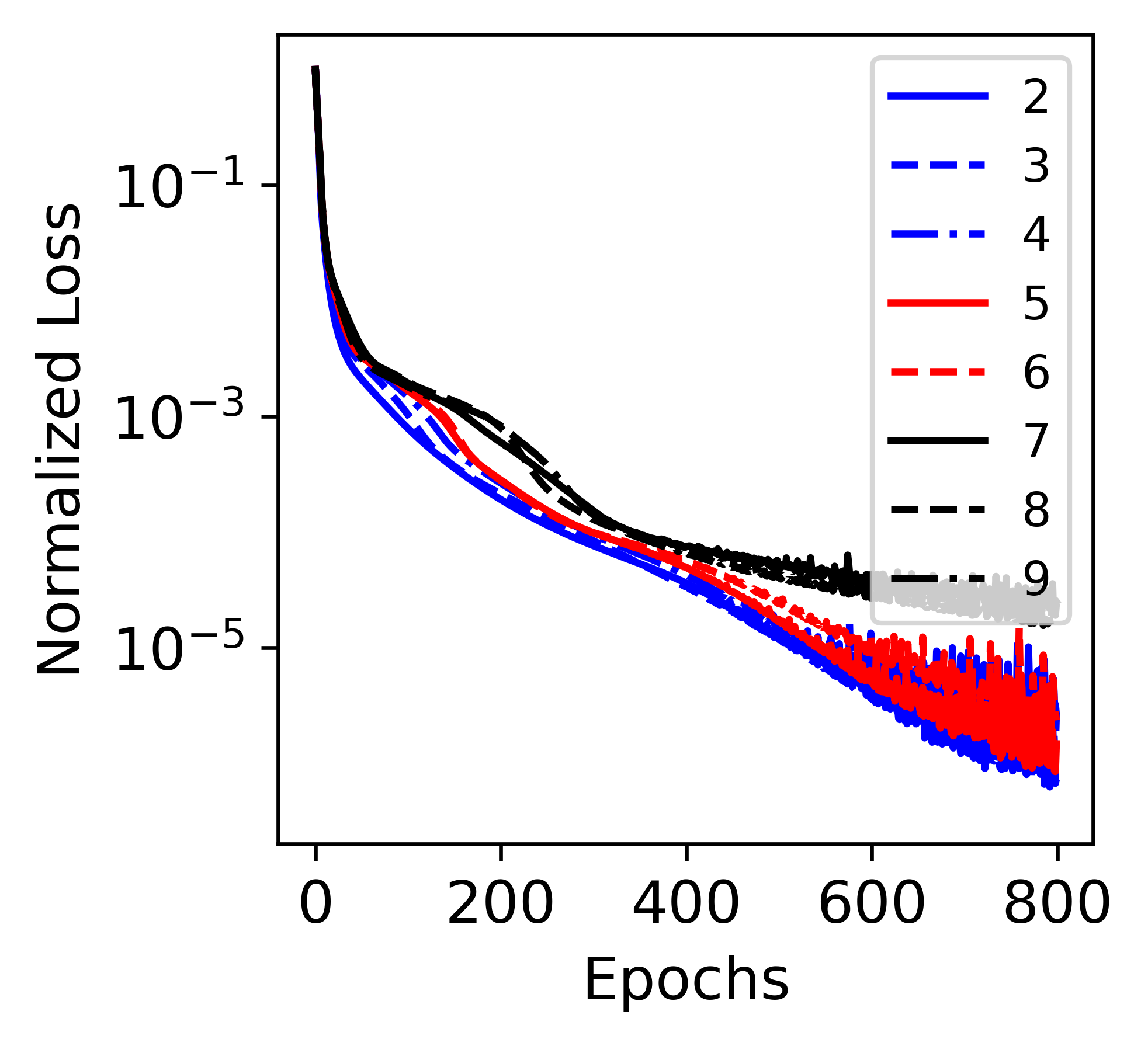}
         \caption{SLO}
         \label{fig_horiz_s}
     \end{subfigure}
     \begin{subfigure}[b]{0.48\linewidth}
         \centering
         \includegraphics[width=\textwidth]{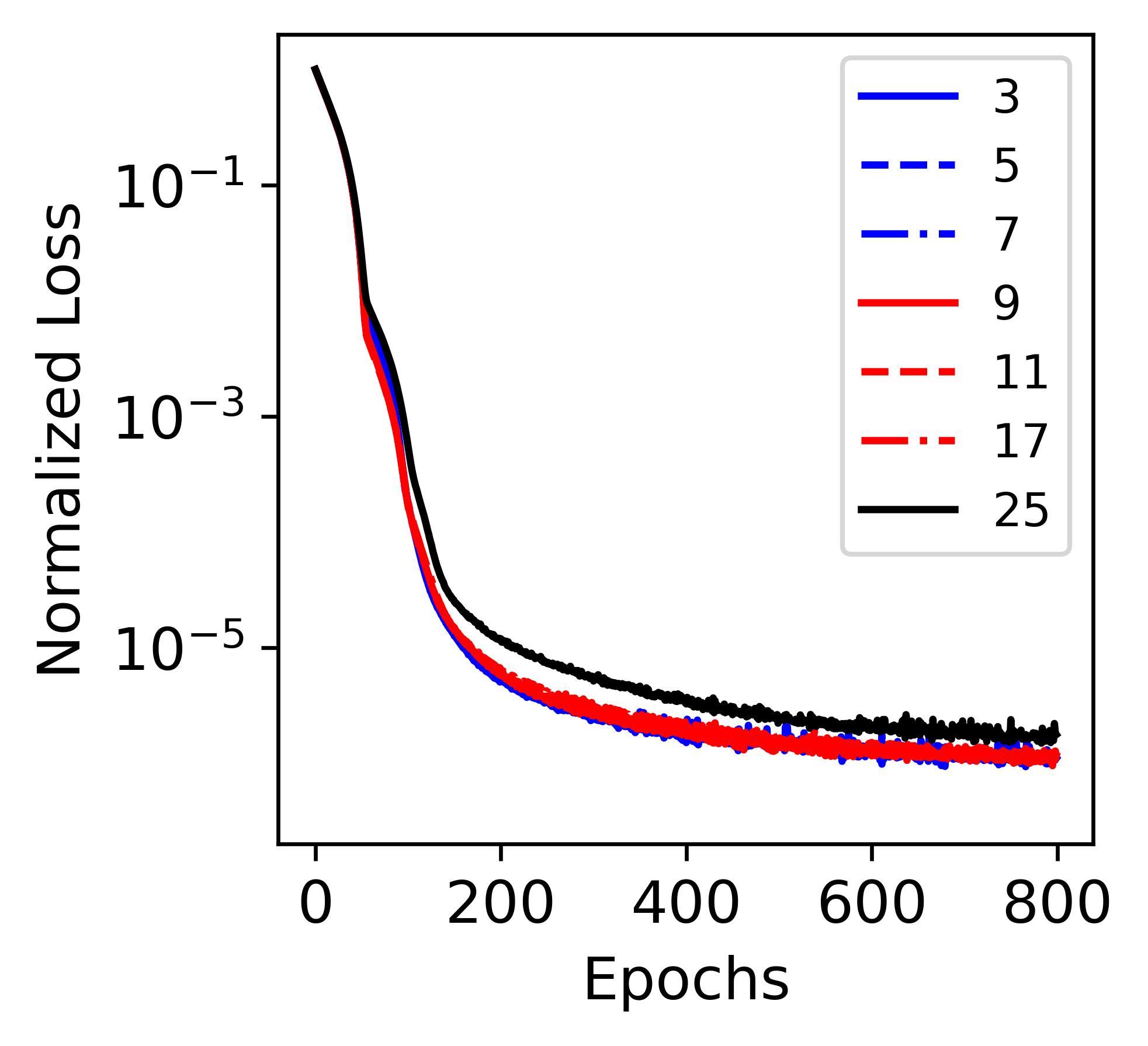}
         \caption{BLO}
         \label{fig_horiz_b}
     \end{subfigure}
        \caption{Effect of horizon length.}
        \label{fig_horiz}
    \vspace{-5mm}
\end{figure}

\begin{figure}
     \centering
     \begin{subfigure}[b]{0.48\linewidth}
         \centering
         \includegraphics[width=\textwidth]{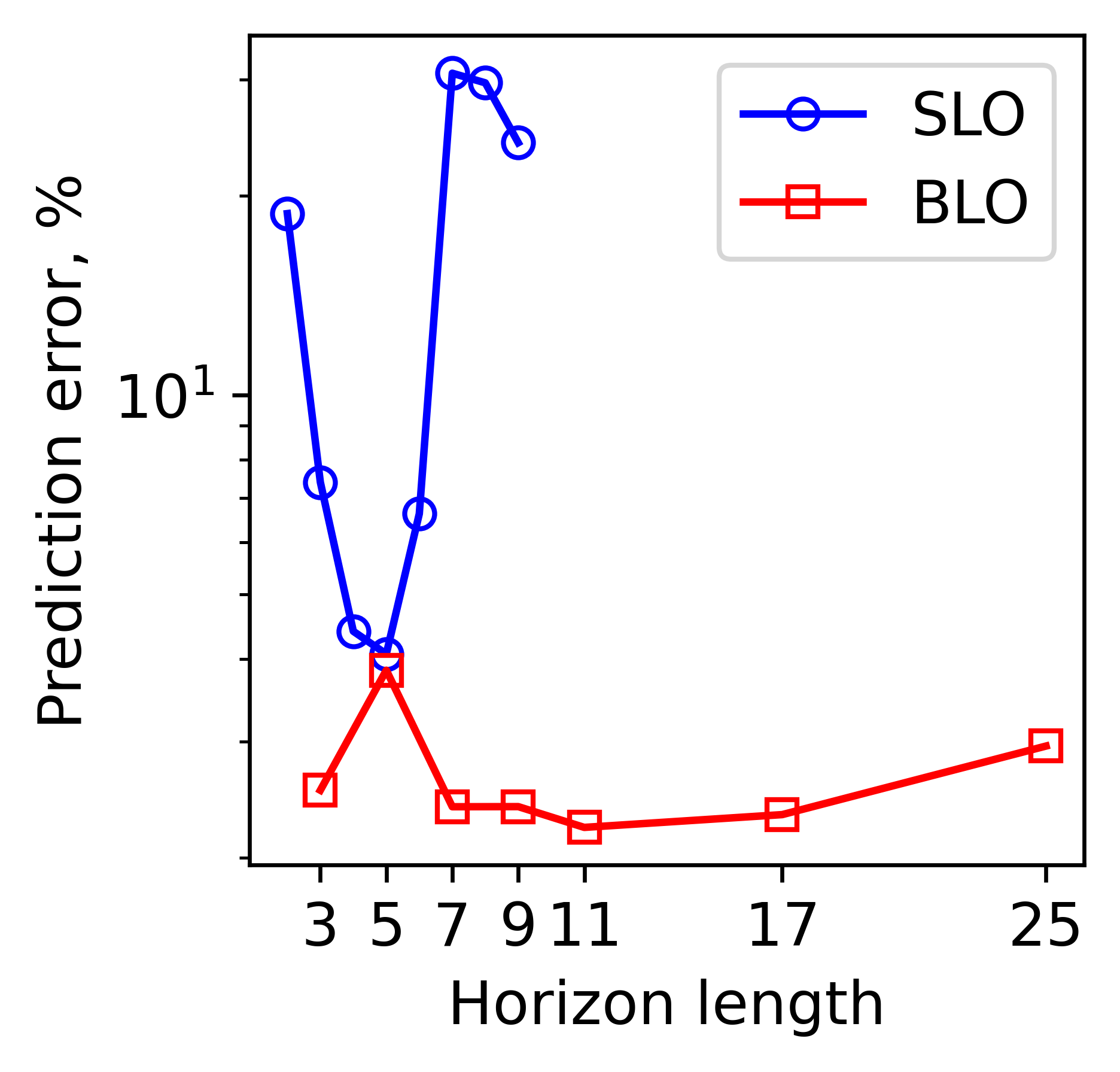}
         \caption{Error in prediction}
         \label{fig_horiz_err}
     \end{subfigure}
     \begin{subfigure}[b]{0.48\linewidth}
         \centering
         \includegraphics[width=\textwidth]{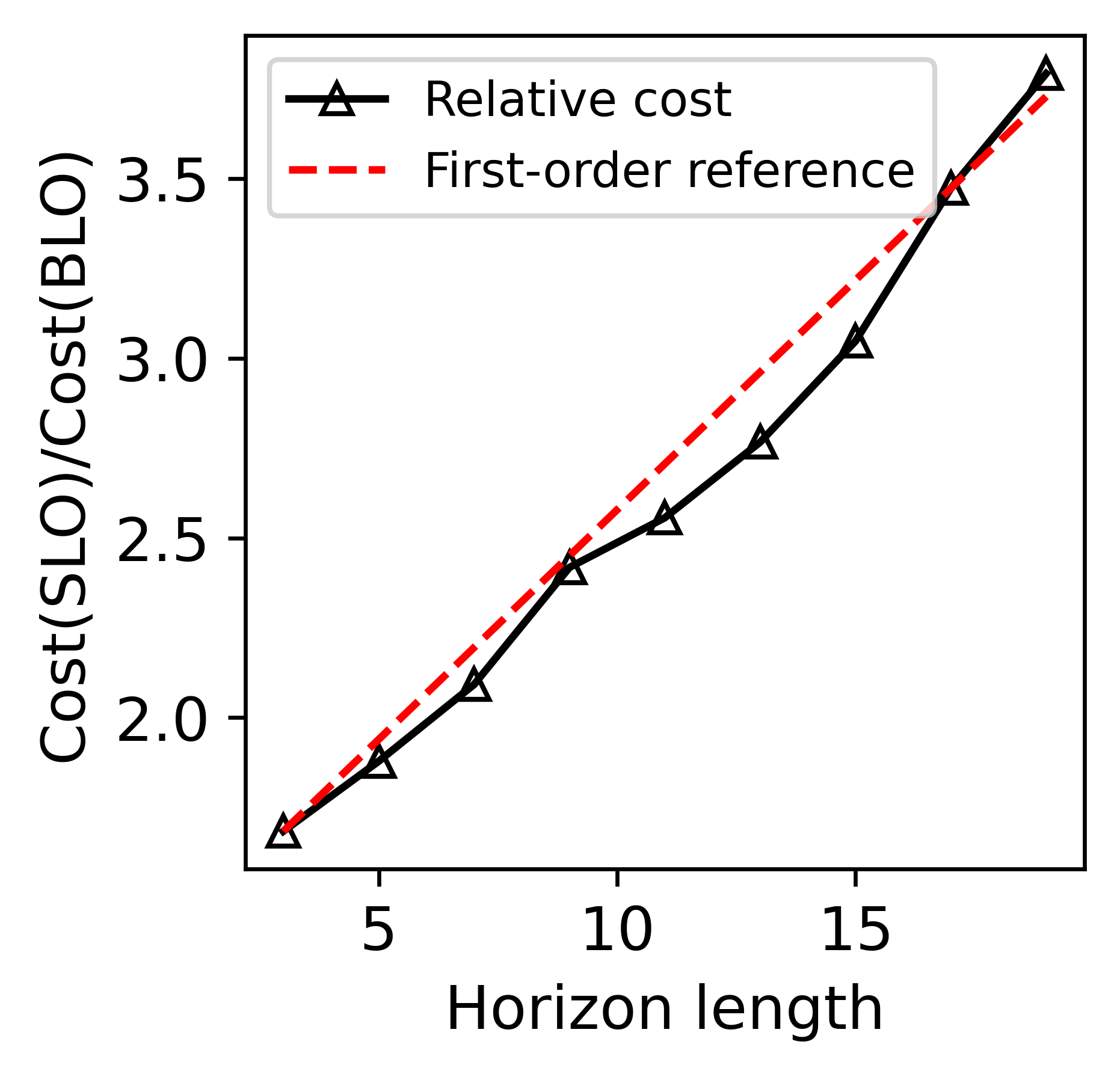}
         \caption{Cost in training}
         \label{fig_horiz_time}
     \end{subfigure}
        \caption{Comparison between BLO and SLO.}
        \label{fig_comp}
    \vspace{-5mm}
\end{figure}

Next, a parametric study is performed on the length of horizon, as shown in Fig. \ref{fig_horiz}.  The SLO starts to show difficulty in convergence when $N>6$, while the BLO consistently achieves high convergence rate up to $N=25$.  The SLO-based training with $N>9$ is not performed due to the high computational cost.
Figure \ref{fig_horiz_err} shows the model prediction error, where both SLO and BLO show a decrease in model error when $N$ is small, however, the error in SLO model quickly increases when $N>6$ while the BLO model error remains consistently low.
Figure \ref{fig_horiz_time} compares the growth in computational cost with increasing horizon length.  For each horizon length, each of the SLO and BLO cases are run for 5 epochs, the time costs are recorded, and the ratio is reported.  It is clear that the complexity of BLO is $O(N)$ less than that of SLO, which is attributed to the removal of the nested formulation from SLO.

\begin{figure}
\begin{center}
\setlength{\abovecaptionskip}{0pt}
\includegraphics[width=1.0\linewidth]{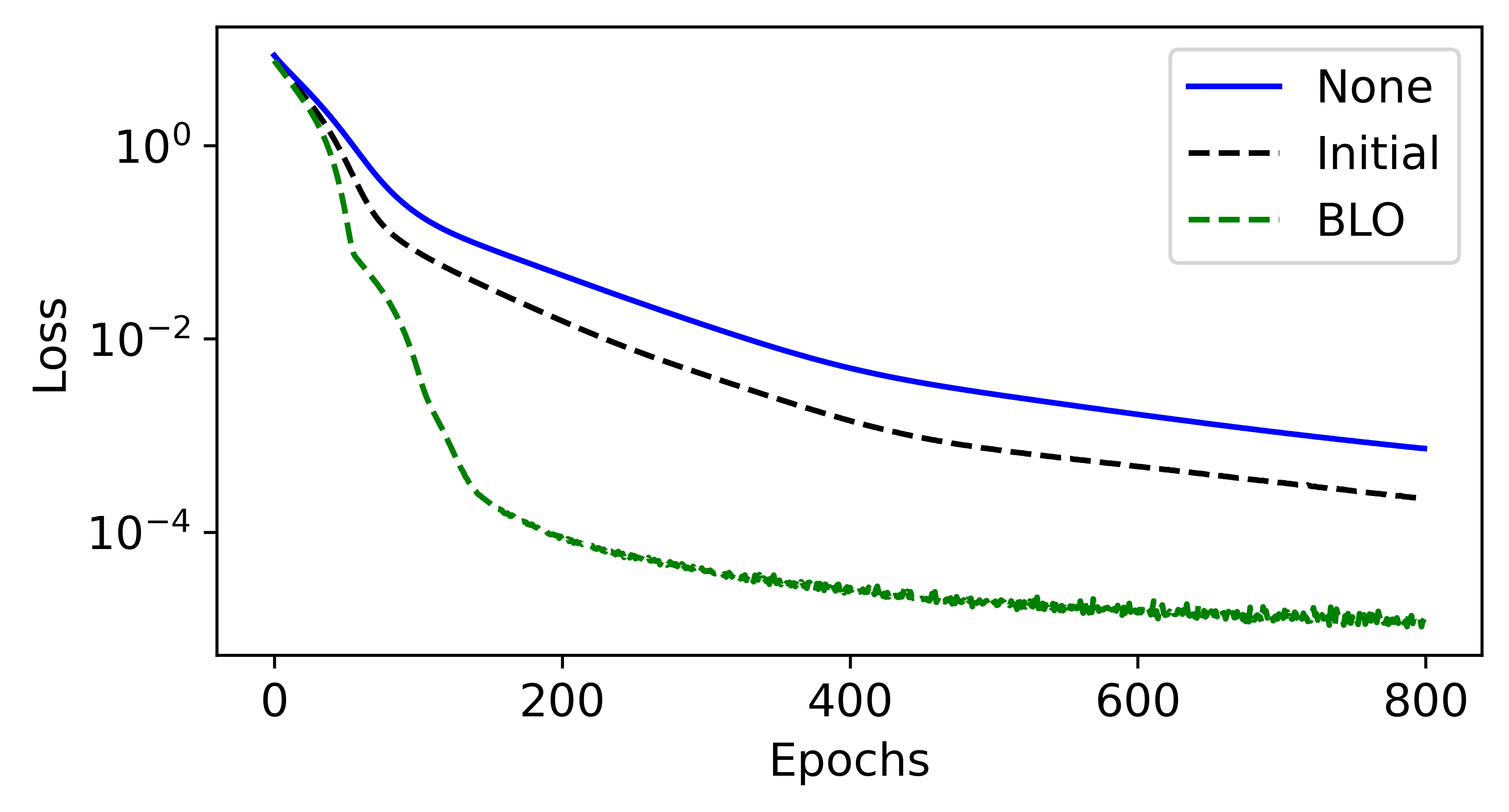}
\end{center}
\vspace{-10pt}
   \caption{Effect of BLO on convergence.}
   \vspace{-0.5cm}
\label{fig_rst}
\end{figure}

Lastly, we also briefly show the effect of bi-level formulation, as shown in Fig. \ref{fig_rst}, where three cases are considered: (1) ``None'': The BLO loss \eqref{eqn_bopt_a} is treated as if SLO and $\vtG,\vtf,\vty$ are optimized together with random initial guesses; (2) ``Initial'': $\vtG$ is computed only once at the start of training and used as an initial guess, and then the ``None'' strategy is used; (3) ``BLO'': The proposed algorithm.  The only case that achieves sufficient convergence is BLO; this is attributed to the optimality of the system matrix maintained by the inner optimization.

\subsection{Double Pendulum}

Next, we consider a damped and controlled double pendulum problem to show the feasibility of the proposed algorithm for high-dimensional systems.
\begin{align}\nonumber
    \ddot{\theta}_1 &= (M_2L_1 \dot{\theta}_1^2 \sin\delta\cos\delta + M_2g\sin\theta_2\cos\delta \\
    &+ M_2L_2\dot{\theta}_2^2\sin\delta - \bar{M}g\sin\theta_1 + u_1)/(L_1\rho) - \dot{\theta}_1 \\\nonumber
    \ddot{\theta}_2 &= (-M_2L_2 \dot{\theta}_2^2 \sin\delta\cos\delta + \bar{M}g\sin\theta_1\cos\delta \\
    &- \bar{M}L_1\dot{\theta}_1^2\sin\delta - \bar{M}g\sin\theta_1 + u_2)/(L_2\rho) - \dot{\theta}_2
\end{align}
where $\delta=\theta_2-\theta_1$, $\bar{M}=M_1+M_2$, and $\rho=\bar{M}-M_2\cos^2\delta$.  The masses and lengths of the two pendulums are $M_1 = 1 \text{kg}$ (upper), $M_2 = 1 \text{kg}$ (lower), $L_1 = 1 \text{m}$, $L_2 = 1 \text{m}$.  The damping terms are added to both pendulums, so as to create a stable isolated equilibrium point in the system.  320 trajectories were generated with initial conditions randomly generated for $\theta_i \in [-10\degree,10\degree]$ and $\dot{\theta}_i \in [-10\degree/s,10\degree/s]$ and control $u_i\in [-0.25, 0.25]\textrm{N}$.  Another 100 trajectories are generated for test.
The embedded space is of dimension 9, with 8 being learned through an autoencoder, and one added to be the constant 1. The encoder and decoder hidden state sizes are (32, 32, 32). all the rest of the details are the same as those for the first example.

\begin{figure}
\begin{center}
\setlength{\abovecaptionskip}{0pt}
\includegraphics[width=1.0\linewidth]{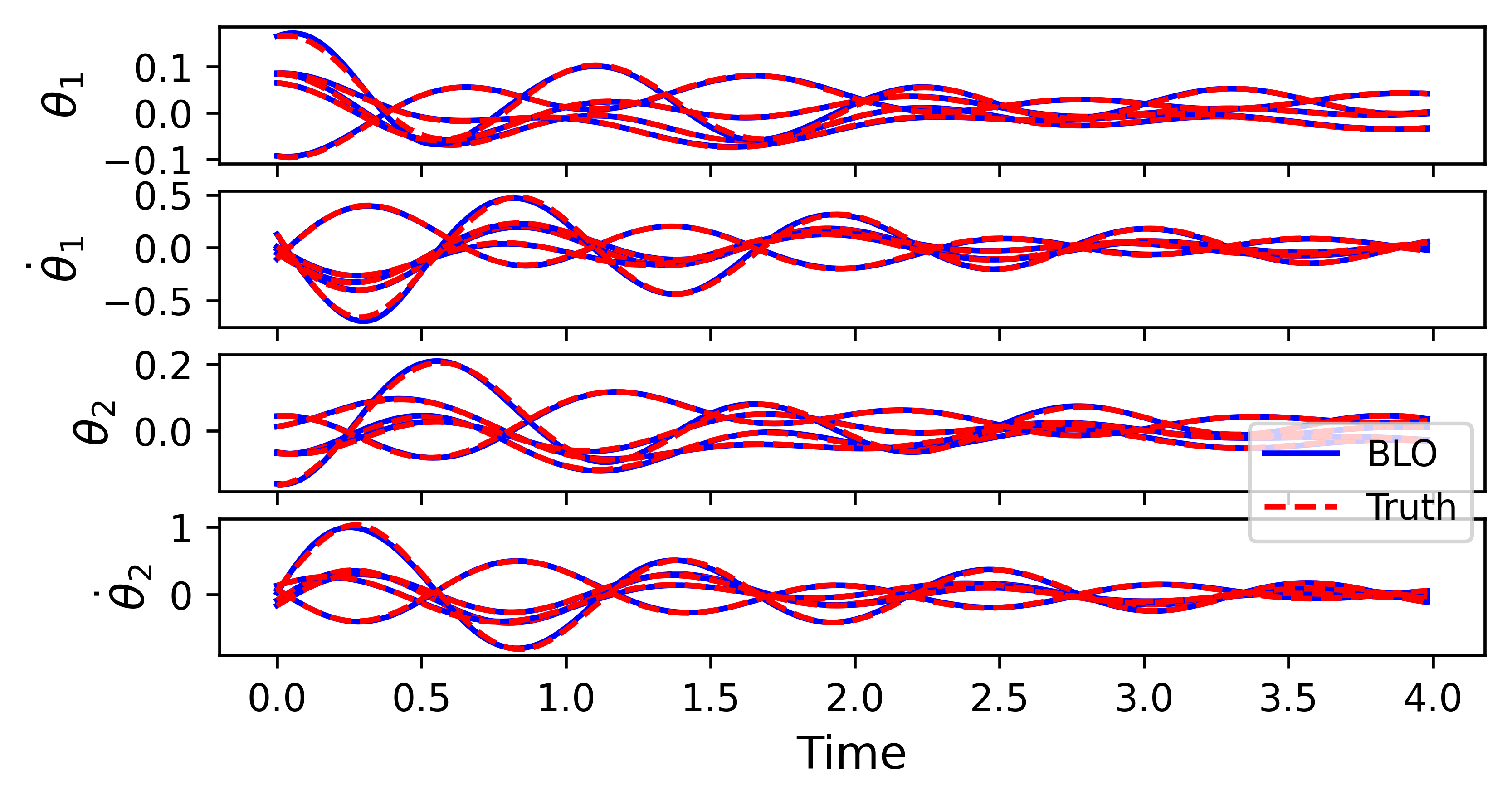}
\end{center}
\vspace{-5mm}
   \caption{Predicted double pendulum dynamics.}
   \vspace{-0.5cm}
\label{fig_dp}
\end{figure}

Figure~\ref{fig_dp} shows the prediction performance of the learned Koopman model and proposed method performs well for this higher dimensional system.  While the training data is sampled at 12.5 Hz for 2s, the prediction is performed at 50 Hz for 4s, thanks to the new continuous-time formulation.  The model matches with the truth well with an error of 4.5\%.  The SLO produces models that have over 50\% error and the predicted trajectories are not shown; the high error is likely due to the low data sampling rate, for which the discrete-time formulation is inaccurate.

\section{Conclusion}

This paper presents a bi-level optimization framework to learn the Koopman Bilinear Form by jointly optimizing the Koopman embedding and dynamics with explicit and exact constraints of continuous-time Koopman dynamics. 
Our approach produces a continuous-time KBF model that is more accurate than the commonly used zeroth-order hold model by construction.  Using an integral formulation, a long-horizon KBF dynamic constraint can be imposed during the learning process without needing to resort to a multi-step discrete-time constraint, that is time consuming to evaluate.  Furthermore, using a bi-level optimization strategy, the KBF dynamics and the nonlinear mapping in a staggered format, and a high convergence rate in training is achieved.

We validate the proposed approach on two example nonlinear systems with control. Results show that our method successfully learns the nonlinear dynamics. Comparing to the single-level optimization method, our method achieves more accurate prediction with lower prediction error, faster convergence, and higher computational efficiency.
Future work will investigate online model prediction for more complex physical scenarios, including the aerial vehicle flying in the wind gust environment. The developed bi-level optimization framework will be released via our open-source robotic learning library PyPose~\cite{wang2023pypose}.

\bibliographystyle{IEEEtran}
\bibliography{references.bib}

\end{document}